\documentstyle[preprint,aps]{revtex}

\begin{document}
\draft
\title{Theory of Neutron Diffraction from the Vortex Lattice in UPt$_3$}
\author{Robert Joynt}
\address{
Department of Physics and Applied Superconductivity Center\\
University of Wisconsin-Madison\\
1150 University Avenue\\
Madison, Wisconsin 53706\\
and\\
Materials Physics Laboratory\\
Helsinki Universty of Technology\\
FIN-02150 Espoo, Finland\\}
\date{\today}
\maketitle
\begin{abstract}
Neutron scattering experiments have recently been performed 
in the superconducting state of UPt$_3$ to determine the 
structure of the vortex lattice.  The data show anomalous 
field dependence of the aspect
ratio of the unit cell in the B phase.  There is apparently
also a change in the effective coherence length on
the transition from the B to the C phases.  
Such observations are not consistent with conventional
superconductvity.  A theory of these results
is constructed based on a picture of two-component superconductivity
for UPt$_3$.  In this way, these unusual observations
can be understood.  There is a possible discrepancy 
between theory and experiment
in the detailed field dependence of the aspect ratio.  
\end{abstract}
\narrowtext

Experiments of the last several years on superconducting UPt$_3$
have revealed a rich phase diagram.  The phase boundaries have
been mapped with considerable accuracy by locating anomalies
in specific heat \cite{fisher} \cite{hasselbach},
ultrasound \cite{qian},
and torsional oscillator signals \cite{kleiman}.
The additional transitions found may be explained by assuming that
the superconductivity is unconventional, and that the order parameter
has two complex components.  \cite{sigrist}
At zero pressure, there are three superconducting phases,
called the A phase (low fields and high temperatures),
the B phase (low field, low temperature),
and the C phase (high field, low temperature).
The theoretical explanation remains untested,
however, unless detailed experimental characterization of the 
phases can be carried out, preferably by searching for definite
signatures of the predicted phases.  The theory suggests
that the B phase is the most
promising in this regard.  In this phase,
the two components are predicted to
coexist, and their interplay can give rise to novel effects.

The comparison of theory and experiment is not straightforward,
however, as no microscopic experimental probe 
couples directly to the order parameter.  
Because of this, it is desirable to have a probe for 
which predictions can be made using 
the phenomenological Ginzburg-Landau theory alone,
thus circumventing microscopic details.
Neutron diffraction from the flux lattice depends only on the
field distribution in the sample \cite{degennes}
and therefore may be calculated
from the phenomenological theory, yet it offers a level of 
detail which cannot be obtained by any other means \cite{bell}.  A new
experiment of this kind has recently been carried out 
and is reported in the accompanying Letter \cite{bell2}.

Calculations of the lattice structure and the 
field distribution have so far been 
limited to the neighborhood of the upper critical field $H_{c2}$
\cite{eplett} \cite{barash} \cite{gc}.
However, it is only at low temperatures that the penetration depth
is short enough that experiments can be
done.  In practice this means 
well away from $H_{c2}$.  Furthermore, 
the interesting effects occur in the 
field dependence of the scattering intensities 
in the B phase, and particularly near $H^*$,
the transition field from the B to the C phase.
The purpose of this Letter is to present
calculations of the lattice structure and the scattering cross section
as a function of applied field, and to compare the results to
experiment.  

The scattering amplitude for a change in
neutron wavevector $\vec{Q}$ in the Born approximation is
\cite {degennes}
\begin{equation}
f_B(\vec{Q}) = \frac{i m_N \mu_N}{\hbar^2(2 \pi |\vec{k}_i|)^{1/2}}
\int d^2r h(\vec{r})e^{i\vec{Q}\cdot \vec{r}}=
\frac{i m_N \mu_N A}{\hbar^2(2 \pi |\vec{k}_i|)^{1/2}}
h(\vec{Q}),
\end{equation}      
where $m_N$ and $\mu_N$ are the mass and magnetic moment of the 
neutron and $\vec{k}_i$ is the incoming 
wavevector.  $h(\vec{r})$ is the
field in the sample and $h(\vec{Q})$ is its Fourier transform.
The integration is over
a cross-section A of the sample perpendicular to the beam.
The object of interest, the observed intensity $I(\vec{Q})$, is thus 
proportional to $|h(\vec{Q})|^2$.
The experiments have all been performed with $\vec{H}$, the
applied field, directed in the basal plane, so the calculations
presented here will be only for that field direction.
I deal only with Bragg scattering, so $\vec{Q}$ is a
reciprocal lattice vector.

If the y-axis (in the basal plane of the hexagonal crystal)
is taken as the field direction,
two-component theories have the generic free energy density:
\begin{eqnarray}
f(\vec{\eta}) &=& \alpha (|\eta_x|^2 + |\eta_y|^2)
+\beta_1 (\vec{\eta}\cdot\vec{\eta}^*)^2
+\beta_2 |\vec{\eta}\cdot\vec{\eta}|^2\\
& & +K_x|D_x\eta_x|^2 + K_y|D_x\eta_y|^2
+K_z(|D_z\eta_x|^2+ |D_z\eta_y|^2).
\label{eq:fe}
\end{eqnarray}
$\eta_x$ and $\eta_y$ are the two components of the order
parameter.  $D_x \equiv -i \partial_x-2 e A_x/ \hbar c$
and similarly for $D_z$.  $\eta_x$ and $\eta_y$  
may belong to one of the two-dimensional representations
$E_1$ or $E_2$ of the point group, or they may 
belong to different, but accidentally nearly degenerate
representations.  The scattering experiments probably
do not distinguish these possibilities.
For simplicity, I have taken $\eta_x$ and $\eta_y$ to be
degenerate (only a single $\alpha$) at zero field.  
This is reasonable
at the low temperature (50 mK) at which the experiments are done.
In fact I shall fix the temperature and treat 
$\alpha$ as constant.  This means that only
deal with the B (low field or $H < H^*$) and C (high field or
$H > H^*$) phases,
and the transition between them, will be discussed.
The free energy 
$F=\int f[\vec{\eta}(\vec{r})] d^3 r$ 
where $f[\vec{\eta}(\vec{r})]$ 
is given by Eq.\ \ref{eq:fe} does indeed lead to the
observed phase diagram with three
superconducting phases for $\vec{H}$ in the basal plane \cite{eplett}.  
The form of $\vec{\eta}(\vec{r})$ in the 
C phase is known.  It is found that $\vec{\eta}(\vec{r}) = 
(\eta_x(\vec{r}),0)$.  Thus the free energy
reduces to
\begin{equation}
f_C(\vec{\eta}) = \alpha |\eta_x|^2
+(\beta_1 +\beta_2) |\eta_x|^4
+K_x|D_x\eta_x|^2 + K_z|D_z\eta_x|^2.
\end{equation}
The problem represented by this free energy density
is isomorphic to that of an s-wave superconductor
with mass anisotropy.  One may therefore transcribe
well-known results \cite{kogan}:
\begin{equation}
|h(\vec{Q})|^2 = H^2 \left[ 1 + (\lambda_z^C)^2 Q_x^2 + (\lambda_x^C)^2 Q_z^2) \right]^{-2}
\exp \left\{ -\frac{1}{2} \left[ (\xi_x^C)^2 Q_x^2 + (\xi_z^C)^2 Q_z^2 \right] \right\}, 
\label{eq:hqc}
\end{equation}
where $\xi_i^C = (K_i / \alpha)^{1/2}$ 
are the coherence lengths and
$\lambda_i^C = (\hbar^2 c^2 / 32 \pi e^2 K_i |\eta_x|^2)^{1/2}$
are the penetration depths in the C phase.  
Since the free energy density
may be transformed to the isotropic form by rescaling
coordinates: $ x' = (\xi_z/\xi_x)^{1/2} x$
and $ z' = (\xi_x/\xi_z)^{1/2} z$, the fluxons
form a rescaled hexagonal lattice (centered rectangular lattice).
In London theory the
vortex cores act as delta-function sources of the field.
Corrections to this require numerical calculations
which have been carried out by Brandt \cite{brandt}, and it is concluded
that the sources are well represented by Gaussians -
this is reason for the exponential factor in Eq.\ 
\ref{eq:hqc}.  The coordinate scaling determines 
the opening angle $\alpha_L$ of the reciprocal lattice (defined in
Fig.\ 1), which uniquely determines the aspect ratio of the unit cell.  
It is given
by $\tan^2(\alpha_L) = 3 K_x / K_z$.  
This result and flux quantization allow us to 
calculate the reciprocal lattice vectors as a function of field.
Substituting these values into Eq.\ \ref{eq:hqc} shows that
$I(\vec{Q})$ is the same for all $\vec{Q}$ in the first shell
(the six smallest nonzero $\vec{Q}$).
These are the only points 
measured to date.

Theory thus predicts that three
properties qualitatively characterize the C phase:\\
(i)~the lattice structure (shape of the unit cell) is 
independent of field;\\
(ii)~all peaks in the first shell have the same intensity;\\
(iii)~the intensities fall off exponentially with field, with
$- d \ln I / dH \sim \xi_x^C \xi_z^C$.\\
Property (i) has been pointed out before \cite{japnum}, \cite{barash}.
The C phase has no unique signatures of unconventional superconductivity, 
however, since an s-wave superconductor with mass anisotropy
has all of these features.

The B phase is quite different.  I first develop the theory and
then turn to comparison with experiment. 

The opening angle in the B phase can be computed in
the regime where $\xi_i << a <<\lambda_i$, where $a$ is the 
lattice constant for the vortices.  In terms of the field,
this is $H_{c1} << H << H_{c2}$.  Since UPt$_3$ is strongly
type-II, this is a substantial range.  It will suffice for our
purposes to compute the currents    
at a distance of order $a$ from the cores,
because the momentum transfers of experimental interest
are of order $1/a$.  Thus the structure of the cores at
short distances of order $\xi_i$, a complex subject,
is not of interest here.  At these larger distances
the $\beta_2$ term in Eq.\ \ref{eq:fe} locks the relative phase 
of $\eta_x$ and $\eta_y$: $\eta_y = \pm i r(H) \eta_x$,
where $r(H)$ is a real ratio.  $r(H)$ supplies the 
interesting field dependence in the B phase.  Since 
$\vec{\eta}_y$ appears continuously, $r(H)$ is
a nonnegative, monotonically decreasing function of $H$ with
$r(H^*) = 0$.  The phase-locking relation, together
with the free energy of Eq.\ \ref{eq:fe}, gives a London
equation for the currents with the penetration depths
\begin{equation}
(\lambda_x^B)^2 = \frac{\hbar^2 c^2}
{32 \pi e^2 |\eta_x|^2 [K_x+r^2(H) K_y]},
\end{equation}
and
\begin{equation}
(\lambda_z^B)^2 = \frac{\hbar^2 c^2}
{32 \pi e^2 |\eta_x|^2 K_z [1+r^2(H)]}.
\end{equation}
In the field regime under consideration, these currents determine
the lattice structure and one finds
\begin{equation}
\tan^2(\alpha_L) = \frac{K_x + r^2(H) K_y}{3 [1+r^2(H)] K_z}. 
\label{eq:bshape}
\end{equation}
{\it The shape 
of the unit cell is strongly field-dependent 
in the B phase.}
This effect does not occur in one-component
superconductors, conventional or unconventional.

The computation of the intensities in the B phase
is not so straightforward,
since corrections to London theory are involved.
However, the same phase-locking approximation
allows us to make a mean-field-type theory.
We neglect amplitude correlations in the region
where the distance from the cores is much greater than
the coherence lengths.  Then the effective field on 
either component has the same spatial dependence 
as in the s-wave case, and we may again apply the
results of Brandt.  The complication which arises is that
both $\vec{\eta}_x$ and $\vec{\eta}_y$ act as sources of the
field.  This leads to separate exponential dependences,
and the breakdown of the simple relationship
$- d \ln I / dH \sim \xi^2$.  There are now two effects which determine
the field dependence of the intensity.
First, the interaction of $\vec{\eta}_x$ and $\vec{\eta}_y$
given by the quartic terms in Eq.\ \ref{eq:fe} sets up
effective fields which give a field dependent 
renormalization of the correlation lengths of both components.
This means that the exponents acquire an additional
field dependence relative to Eq.\ \ref{eq:hqc}.
Second, the prefactors of the exponents have a 
field dependence because of the separate contributions from
$\vec{\eta}_x$ and $\vec{\eta}_y$,
whose relative weight is field-dependent.

The full expression for the $h(\vec{Q})$ 
may be separated into a part which depends relatively
weakly on $H$ and is probably unobservable,
and the exponential part \cite{kleiman}.  The full expression, and details of its
derivation, will be given elsewhere.  The exponential part is
\begin{eqnarray}
h(\vec{Q}) & \sim & [(1 + r^2)^{-1}
F(\frac{\xi^B_{xx}}{\xi^B_{z}}, \xi^B_{xx}Q_x,\xi^B_{zx} Q_z)\\
&&+(1+r^2 K_y / K_x)^{-1} F(\frac{\xi^B_{z}}{\xi^B_{xx}},
\xi^B_{zx} Q_z,\xi^B_{xx}Q_x)]
   \times \exp[(-(\xi^B_{xx})^2 Q_x^2 - (\xi^B_{zx})^2 Q_z^2)/4] \\
&& + [(1 + r^{-2})^{-1}
F(\frac{\xi^B_{xy}}{\xi^B_{zy}}, \xi^B_{xy}Q_x,\xi^B_{zy} Q_z)\\
&&+(r^{-2} + K_x / K_y)^{-1} F(\frac{\xi^B_{zy}}{\xi^B_{xy}},
\xi^B_{zy} Q_z,\xi^B_{xy}Q_x)]
   \times \exp[(-(\xi^B_{xy})^2 Q_x^2 - (\xi^B_{zy})^2 Q_z^2)/4].  
\label{eq:bint}
\end{eqnarray}
In this formula, the coherence length $\xi^B_{xx}$ is given
by
\begin{equation}
\xi^B_{xx} = (\frac{K_x}{\alpha})^{1/2}
\left\{ 1 + \frac{1}{4}[(1-\frac{\beta_2}{\beta_1})
(2 - \frac{H}{H_{c2}}-\frac{H}{H_y})]
-\frac{1}{4}[(1-\frac{\beta_1}{\beta_2})
(\frac{H}{H_{c2}}-\frac{H}{H_y})] \right\}^{1/2},
\end{equation}
and there are similar formulas for the other three coherence lengths.
The field $H_y$ is a constant given by
\begin{equation}
H_y = \frac{H^*}{1 - b + b H^*/H_{c2}},
\end{equation}
with $b \equiv (\beta_1-\beta_2)/(\beta_1+\beta_2)$,
while the function $F(x, q_1, q_2)$ is defined as
\begin{equation}
F(x, q_1, q_2) = x \int_{-\infty}^{-\infty} du~
\int_{-\infty}^{-\infty} dv~
\frac{u^2}{x^2u^2 + v^2}
\exp[-(u-i q_1 / 2)^2 - ( v-i q_2 / 2)^2].
\end{equation}

Thus the effective coherence length, if it is defined by the
slope of $\ln I(\vec{Q})$ as a function of field, is seen to
be field dependent.  This again is utterly characteristic
of a multicomponent system, and cannot be found in a one-component
superconductor.  

These calculations have been done assuming that the
separation of the singularities in 
$\vec{\eta}_x$ and $\vec{\eta}_y$ in the unit
cell are separated by a distance much less than the lattice
constant.  If this is not the case, as in the 
'shifted' phase predicted near the tetracritical point, then the intensities
for the reciprocal lattice vectors in the first shell
may be quite different because of extinction effects.
In particular, at low field, the intensity for $\vec{Q}_{0,1}$
will be less than that for $\vec{Q}_{1,-1}$.
Unfortunately, explicit calculations for this phase 
are quite difficult.

Eqs.\ \ref{eq:bshape} and \ref{eq:bint}
give three qualitative predictions for the B phase:\\
(i)~the lattice structure (shape of the unit cell) is 
depends on field;\\
(ii)~peaks in the first shell have different intensities;\\
(iii)~the intensities have a complex field dependence, 
with deviations from the exponential shape; \\  
(iv)~there is a kink in I(H) at the second-order
B-C transition. 
Property (i) has been pointed out before \cite{eplett}, \cite{japnum}, \cite{barash}.
Property (iv) does {\it not} imply that the 
BC transition is first-order.  The slope of I(H) is not interpreted as
proportional to a single coherence length, (which would then
be discontinuous).  Instead, the kink is interpreted as 
signalling the {\it continuous} growth of a new component of the scattering.
The BC transition is second-order in this theory.

Before comparing these predictions quantitatively with experiment,
we must discuss the determination of the Ginzburg-Landau 
parameters.  The final results for the opening angle
of the flux lattice depend only on
the ratio $\beta_1/\beta_2$ and the 
ratios of the stiffness constants $K_x, K_y$ and $K_z$.
I take $\beta_1/\beta_2 = 0.5$, as determined by specific 
heat experiments.
The stiffness constant ratios as
determined by fitting to the neutron data
are $K_x:K_y:K_z = 1.5:0.88:2.5$.  

The opening angle is plotted in Fig.\ 1.
The theory predicts a field-dependent lattice
structure in the B phase: $H < H^* = 5.3 kOe$,
as is observed.  There should be a kink in the curve at
$H^*$ and in the high-field C phase, $\alpha_L$ should be
independent of field.  These predictions are
consistent with the data, but more points at higher fields
and smaller error bars are needed to provide a real test.

The logarithm of the form factor $H_1$ \cite{kleiman} at the Bragg point
$\vec{Q}_{1,1}$ 
is plotted as a function of field in Fig.\ 2 with the same parameter ratios
$K_x:K_y:K_z = 1.5:0.88:2.5$.  The coherence length, defined as the 
geometric mean of the coherence lengths in the $x$ and $z$
directions, was taken to be $101 \AA$.
Very good agreement with experiment is obtained.
In particular, the truly novel feature in the
data, the kink at $H = H^* = 5.3 kOe$,
is very well reproduced.
I have also calculated the intensity at the Bragg point $\vec{Q}_{0,1}$.
It is not shown because for the parameter range here the calculated intensities 
differ by only a few per cent.  It will be difficult to 
verify that this difference exists at the current 
level of experimental accuracy. 

The stiffness constant ratios obtained by the present fit
can be compared to ratios obtained by fitting to 
the measured values of the upper
critical fields at temperatures near T$_c$
for different field directions and from the discontinuity in slope at the
tetracritical point.  This gives $K_x:K_y = 2$,
and applying constraints from particle-hole symmetry
yields $K_x:K_y:K_z = 1:2:4$.  It is quite clear that
the two methods disagree, and even the ordering differs
between $K_x$ and $K_y$.  Indeed, if $K_x < K_y$,
as suggested by the critical field slopes,
the curve $\alpha_L(H)$ is monotonically {\it decreasing} for $H < H^*$,
in clear contradiction to the data of Fig.\ (1).  
The significance of the discrepancy is not clear at present,
since it involves
an extrapolation from high to low
temperatures which may not be justifiable.
Nonlocal corrections at low temperatures are surely important
and the nature of these corrections for 
unconventional superconductors is not known.
 
The qualitatively new features predicted for neutron scattering
from the flux lattice in the B phase are observed,
as is the relatively conventional behavior of the C phase.
Both the opening-angle data and the field-dependent intensities
can be fit to good accuracy.  This fit does not agree
with one obtained from critical field measurements.  
This may be due to difficulties of
extrapolation to the low temperatures of the experiment,
or may indicate a real discrepancy between theory and experiment.

This work was supported by the 
National Science Foundation through Grant DMR-9214739,
and by NORDITA, Copenhagen, Denmark.

\begin{figure}
\caption{Theoretical curve and experimental data points
from Ref.\ 8 for the opening 
angle of the flux lattice as a function of the applied field.
The inset shows the definition of the angle.}
\label{fig:angle}
\end{figure}

\begin{figure}
\caption{Theoretical curve and experimental data points
from Ref.\ 8 for the logarithm of the form factor at 
the first Bragg peak as a function of the applied field.}
\label{fig:intens}
\end{figure}

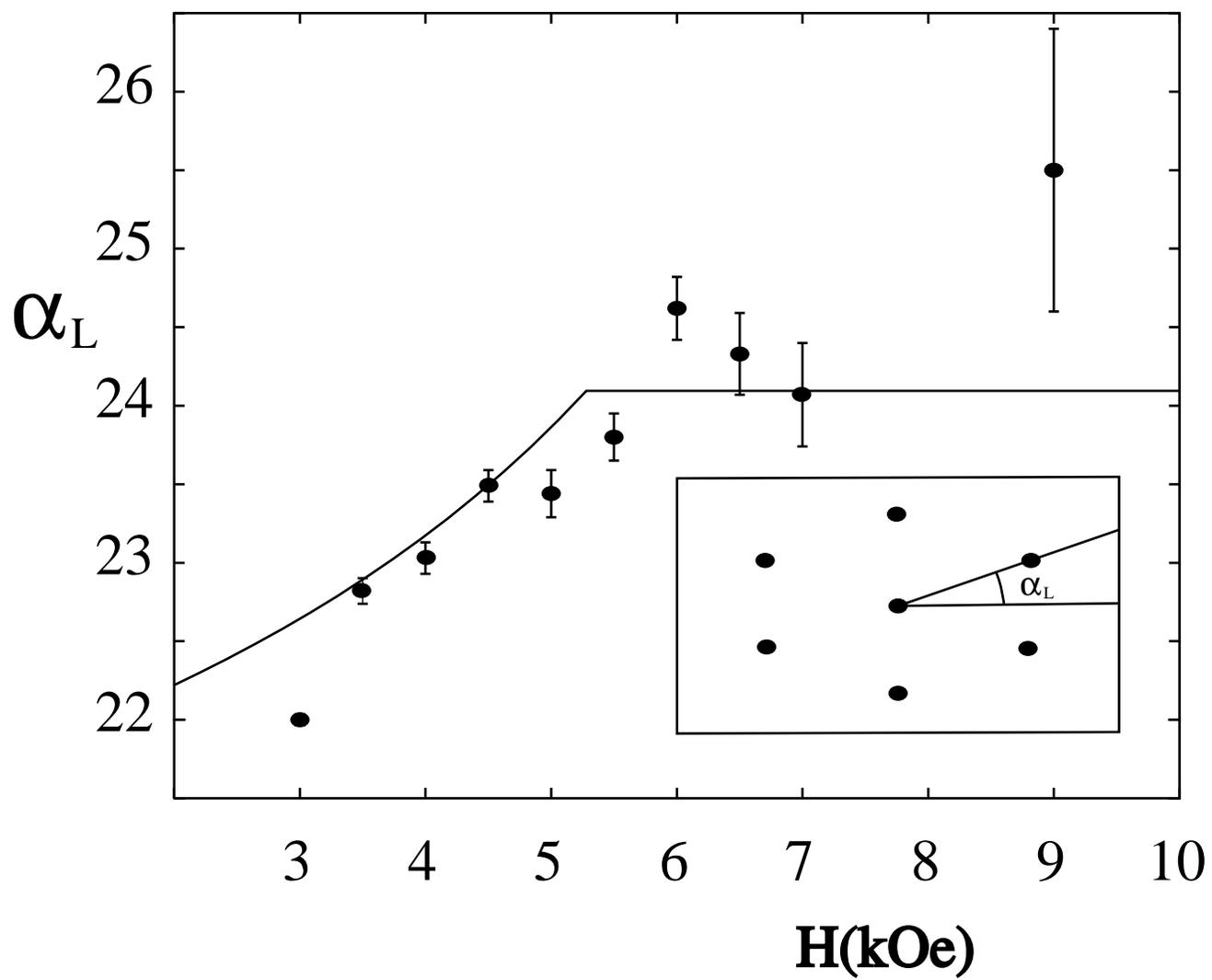
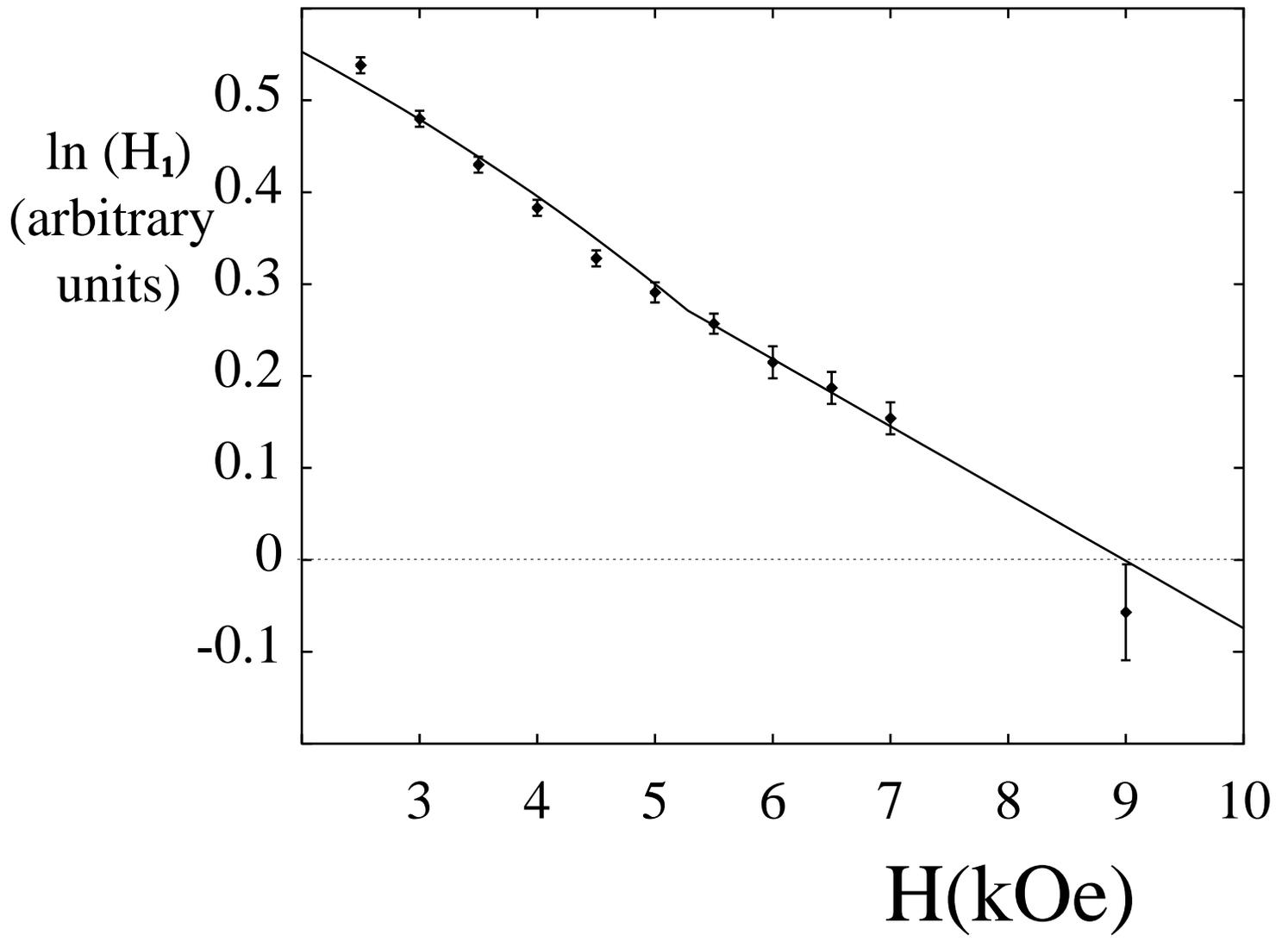
\end{document}